\documentclass[us letters]{article}
\usepackage{amsmath,graphicx,amssymb,delarray,subfigure,verbatim, bbding}
\usepackage{diagbox, makecell, mathtools}
\usepackage{bbding}
\usepackage{multirow}
\usepackage{mathrsfs}
\usepackage{booktabs}
\usepackage{INTERSPEECH2022}

\makeatletter
\renewcommand\normalsize{%
	\abovedisplayskip 3\p@ \@plus3\p@ \@minus3\p@
	\abovedisplayshortskip \z@ \@plus3\p@
	\belowdisplayshortskip 6\p@ \@plus3\p@ \@minus3\p@
	\belowdisplayskip \abovedisplayskip
	\let\@listi\@listI}
\makeatother

\title{MDNet: Learning Monaural Speech Enhancement from Deep Prior Gradient}
\name{Andong Li$^{1, 2}$, Chengshi Zheng$^{1,2}$, Ziyang Zhang$^{3}$, Xiaodong Li$^{1,2}$}
\address{
	$^1$Key Laboratory of Noise and Vibration Research, Institute of Acoustics, Chinese Academy of Sciences, Beijing, China\\
	$^2$University of Chinese Academy of Sciences, Beijing, China\\
	$^3$Advanced Computing and Storage Lab, Huawei Technologies Co.~Ltd., Beijing, China
}
\email{ \{liandong, cszheng, lxd\}@mail.ioa.ac.cn, \{zhangziyang11\}@huawei.com}
\begin{document}
\maketitle
\begin{abstract}
While traditional statistical signal processing model-based methods can derive the optimal estimators relying on specific statistical assumptions, current learning-based methods further promote the performance upper bound via deep neural networks but at the expense of high encapsulation and lack adequate interpretability. Standing upon the intersection between traditional model-based methods and learning-based methods, we propose a model-driven approach based on the maximum a posteriori (MAP) framework, termed as MDNet, for single-channel speech enhancement. Specifically, the original problem is formulated into the joint posterior estimation \emph{w.r.t.} speech and noise components. Different from the manual assumption toward the prior terms, we propose to model the prior distribution via networks and thus can learn from training data. The framework takes the unfolding structure and in each step, the target parameters can be progressively estimated through explicit gradient descent operations. Besides, another network serves as the fusion module to further refine the previous speech estimation. The experiments are conducted on the WSJ0-SI84 and Interspeech2020 DNS-Challenge datasets, and quantitative results show that the proposed approach outshines previous state-of-the-art baselines.

\end{abstract}
\vspace{-0.2cm}
\section{Introduction}
\label{sec:introduction}
\vspace{-0.1cm}
The enhancement of noisy speech in real acoustic scenarios is a challenging task, especially for low signal-to-noise ratios (SNRs) or non-stationary noises~{\cite{loizou2013speech}}. Recently, the advent of deep neural networks (DNNs) has significantly promoted the development of the speech enhancement (SE)~{\cite{wang2018supervised}} and leads to giant performance leaps over traditional methods.

For conventional DNN-based methods, sophisticated network structures are often devised in an end-to-end manner to learn the nonlinear mapping relations between input and output pairs~{\cite{tan2019learning, luo2019conv}}. Despite being feasible and efficient, they lack interpretability as the whole network is designed in a black-box manner. As a solution, some more recent works attempt to decompose the original task into multiple sub-steps and provide intermediate supervision as the prior information to boost the subsequent optimization progressively~{\cite{li2020speech, li2021simultaneous, hao2020masking}}. Increasing results show that compared with the single-stage paradigm with blind prior, the introduction of pre-estimated prior can lead to more accurate target estimation. Nonetheless, it is still far from affirmative how to decompose the mapping task optimally, and current multi-stage strategies seem to be empirical and intuitive.

In traditional statistical signal processing based SE methods, they usually derive optimal complex-spectral or spectral magnitude estimators~{\cite{gerkmann2014bayesian, gerkmann2012mmse}} following specific optimization criteria, \emph{e.g.}, maximum likelihood (ML), the Bayesian maximum a  posteriori (MAP), and minimum squared error (MMSE). In detail, when specific prior terms and conditional distribution assumptions are provided, the optimal parameter estimator can be obtained using Bayes' theorem. It is evident that the performance of these model-based methods largely hinges on the model accuracy and rationality of the prior distribution that is often manually designed and can heavily degrade in adverse environments. In contrast, existing learning (NN)-based methods are highly-encapsulated, which skip the prior estimation and directly map to the target in a data-driven manner. Therefore, it will be attractive and meaningful to investigate the integration of both categories and leverage their respective merits.

In this paper, we propose a model-driven network, termed as \textbf{MDNet}, for monaural speech enhancement. Different from previous blind NN-based works, we devise our method following the MAP criterion, which empowers each module with better interpretability. Concretely, under the MAP guidance, our enhancement task is converted into the joint posterior estimation problem \emph{w.r.t.} speech and noise. Instead of manually designing the prior distribution in traditional SE algorithms, we propose to learn the speech/noise priors from training data, which can effectively fit real parameter distribution. Besides, the unfolding strategy is proposed, where we explicitly predict the prior gradient and update the target in each iterative step. To our best knowledge, this is the first time to propose the deep prior gradient method in the speech front-end field and we expect it to promote the combination of model-based and learning-based methods. We conduct the experiments on the WSJ0-SI84 and DNS-Challenge datasets and quantitative results show that the proposed approach yields competitive performance over current top-performed SE systems.

The rest of the paper is organized as follows. In Section~{\ref{sec:problem-formulation-and-map}}, we formulate the problem and introduce the MAP criterion. In Section~{\ref{sec:proposed-approach}}, the proposed approach is presented. Experimental setup is given in Section~{\ref{sec:experiments-setup}}, and we present the results and analysis in Section~{\ref{sec:results-and-analysis}}. Some conclusions are drawn in Section~{\ref{sec:conclusion}}.
\vspace{-0.60cm}
\section{Problem formulation and MAP}
\label{sec:problem-formulation-and-map}
\vspace{-0.1cm}
In the short-time Fourier transform (STFT) doamin, the observed mixture speech can be modeled as
\begin{gather}
\label{eq2}
X_{k, l} = S_{k, l} + N_{k, l},
\end{gather}
where $\left\{X_{k, l}, S_{k, l}, N_{k, l}\right\}$ are the corresponding variables in the freqency index of $k\in\left\{1,...,K\right\}$ and time index of $l\in\left\{1,...,L\right\}$.

In conventional DNN-based methods, the network serves as the mapping function to estimate the target from input mixture, given as
\begin{gather}
\label{eq3}
\widetilde{\mathbf{S}} = \mathcal{F}\left(\mathbf{X}; \mathbf{\Theta}\right),
\end{gather}
where $\mathcal{F}\left(\cdot;\mathbf{\Theta}\right)$ denotes the network function with parameter set $\mathbf{\Theta}$, and tilde symbol denotes the estimated variable. However, as the network directly estimates the posterior probability from mixtures, it lacks the prior term and the performance may suffer from heavy degradation under low SNRs. To resolve this problem, we reformulate the problem based on the MAP framework:
\begin{gather}
\label{eq4}
\resizebox{0.99\linewidth}{!}{$
\mathop{\arg\max}_{\mathbf{S}, \mathbf{N}} P\left(\mathbf{S}, \mathbf{N} | \mathbf{X}\right) \propto \mathop{\arg\max}_{\mathbf{S}, \mathbf{N}} P\left(\mathbf{X}| \mathbf{S}, \mathbf{N}\right) P\left(\mathbf{S}\right)P\left(\mathbf{N}\right),
$}
\end{gather}
where $P\left(\mathbf{S}, \mathbf{N}|\mathbf{X}\right)$ denotes the joint posterior probability of $\left\{\mathbf{S}, \mathbf{N}\right\}$, $P\left(\mathbf{X}| \mathbf{S}, \mathbf{N}\right)$ is the conditional probability of $\mathbf{X}$, and $P\left(\mathbf{S}\right)$, $P\left(\mathbf{N}\right)$ denote the prior probability of speech and noise. Eqn.({\ref{eq4}}) holds when speech and noise are assumed to be statistically independent.

In traditional SE methods, speech and noise are often assumed to follow certain probability distributions, where complex Gaussian distribution is most widely used~{\cite{ephraim1984speech}}. In contrast, in this study we focus on the reconstruction error $\mathbf{E} = \mathbf{X}-\mathbf{S}-\mathbf{N}$ and assume that it follows the zero-mean multivariate complex Gaussian probability density function (PDF), \emph{i.e.}, $\mathcal{N}_{\mathbb{C}}\left(\mathbf{0}, \mathbf{\Lambda}\right)$. For modeling convenience, we assume $\mathbf{\Lambda}$ is time-invariant and take a negative logarithmic operation on both sides of Eqn.({\ref{eq4}}), it can then be rewritten as
\begin{gather}
\label{eq5}
\mathop{\arg\min}_{\mathbf{S}, \mathbf{N}}\left\|\mathbf{X} - \mathbf{S} - \mathbf{N}\right\|_{F}^{2} + \alpha_{S}\Psi_{S}\left(\mathbf{S}\right) + \alpha_{N}\Psi_{N}\left(\mathbf{N}\right),
\end{gather}
where $\left\{\Psi_{S}\left(\mathbf{S}\right), \Psi_{N}\left(\mathbf{N}\right) \right\}$ are prior terms of speech and noise with distribution parameters $\left\{\alpha_{S}, \alpha_{N}\right\}$. In~{\cite{wang2021compensation}}, the authors revealed the optimization compensation effect between magnitude and phase during the complex spectrum recovery process. To dampen this effect, a collaborative complex spectrum reconstruction method was developed, where the complex spectrum recovery can be decoupled into magnitude filtering with range of $\left(0, 1\right)$ and complex residual mapping~{\cite{li2022glance}}. As such, we rewrite the speech and noise as
\begin{gather}
\label{eq6}
\mathbf{S} = \mathbf{G}_{S}\mathbf{X} + \mathbf{R}_{S},\\
\mathbf{N} = \mathbf{G}_{N}\mathbf{X} + \mathbf{R}_{N},
\end{gather}
where $\mathbf{G}$ and $\mathbf{R}$ namely refer to real-valued gains and complex residual components. If we regard the prior term $\left\{\mathbf{S}, \mathbf{N}\right\}$ as the joint prior of $\left\{\mathbf{G},\mathbf{R}\right\}$ and further assume that they are statistically independent, Eqn.({\ref{eq5}}) can be rewritten as
\begin{equation}
\label{eq7}
\begin{split}
\mathop{\arg\min}_{\mathbf{G}_{S}, \mathbf{G}_{N}, \mathbf{R}_{S}, \mathbf{R}_{S}}&\left\|(\mathbf{1}-\mathbf{G}_{S}-\mathbf{G}_{N})\mathbf{X} - \mathbf{R}_{S} - \mathbf{R}_{N}\right\|_{F}^{2} +\\
 &\alpha_{G_{S}}\Psi_{G_{S}}\left(\mathbf{G}_{S}\right) + \alpha_{G_{N}}\Psi_{G_{N}}\left(\mathbf{G}_{N}\right) \\
 &+\alpha_{R_{S}}\Psi_{R_{S}}\left(\mathbf{R}_{S}\right)+\alpha_{R_{N}}\Psi_{R_{N}}\left(\mathbf{R}_{N}\right).
\end{split}
\end{equation}
\begin{figure*}[t]
	\centering
	\centerline{\includegraphics[width=1.75\columnwidth]{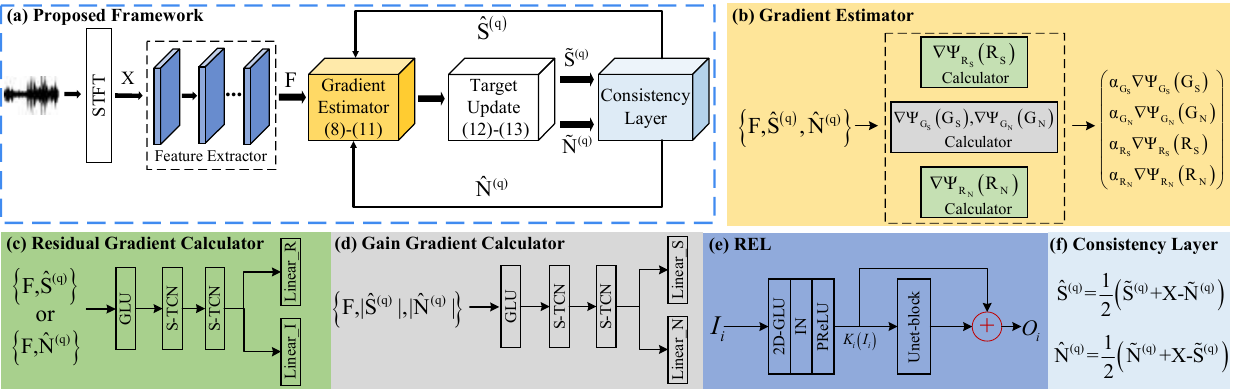}}
	\caption{(a) Overall diagram of the proposed framework. (b) Internal structure of the gradient estimator. (c) Internal structure of the residual gradient calculator. (d) Internal structure of the gain gradient calculator. (e) Internal structure of the recalibration encoding layer. (f) Internal operation of the consistency layer. Different modules are indicated with different colors for better visualization.}
	\label{fig:architecture}
	\vspace{-0.6cm}
\end{figure*}
\vspace{-0.35cm}
\section{Proposed approach}
\label{sec:proposed-approach}
\vspace{-0.1cm}
\subsection{Iterative gradient descent optimization}
\label{sec:iteractive-gradient-descent-optimization}
\vspace{-0.1cm}
To solve the multi-target optimization problem in Eqn.(\ref{eq7}), let us assume the prior terms are differentiable, then the problem can be addressed via the gradient descent method (GDM). Specifically, in the $(q+1)$th iteration, we update the above four targets as follows:
\begin{gather}
\label{eq8}
\resizebox{0.99\linewidth}{!}{$
\widetilde{\mathbf{G}}_{S}^{(q+1)} = \widetilde{\mathbf{G}}_{S}^{(q)} - \eta_{G_{S}}\left(\nabla_{G_{S}}\mathbf{T}^{(q)} + \alpha_{G_{S}}\nabla_{G_{S}}\Psi_{G_{S}}\left(\widetilde{\mathbf{G}}_{S}^{(q)}\right)\right),
$}
\end{gather}
\begin{equation}
\label{eq9}
\resizebox{0.99\linewidth}{!}{$
\widetilde{\mathbf{G}}_{N}^{(q+1)} = \widetilde{\mathbf{G}}_{N}^{(q)} - \eta_{G_{N}}\left(\nabla_{G_{N}}\mathbf{T}^{(q)} + \alpha_{G_{N}}\nabla_{G_{N}}\Psi_{G_{N}}\left(\widetilde{\mathbf{G}}_{N}^{(q)}\right)\right),
$}
\end{equation}
\begin{equation}
\label{eq10}
\resizebox{0.99\linewidth}{!}{$
\widetilde{\mathbf{R}}_{S}^{(q+1)} = \widetilde{\mathbf{R}}_{S}^{(q)} - \eta_{R_{S}}\left(\nabla_{R_{S}}\mathbf{T}^{(q)} + \alpha_{R_{S}}\nabla_{R_{S}}\Psi_{R_{S}}\left(\widetilde{\mathbf{R}}_{S}^{(q)}\right)\right),\\
$}
\end{equation}
\begin{equation}
\label{eq11}
\resizebox{0.99\linewidth}{!}{$
\widetilde{\mathbf{R}}_{N}^{(q+1)} = \widetilde{\mathbf{R}}_{N}^{(q)} - \eta_{R_{N}}\left(\nabla_{R_{N}}\mathbf{T}^{(q)} + \alpha_{R_{N}}\nabla_{R_{N}}\Psi_{R_{N}}\left(\widetilde{\mathbf{R}}_{N}^{(q)}\right)\right),
$}
\end{equation}
where $\mathbf{\eta} = \left\{\eta_{G_{S}}, \eta_{G_{N}}, \eta_{R_{S}}, \eta_{R_{N}}\right\}$ denote the optimization step of the four parameters, and $\mathbf{T}^{(q)}$ denotes the quadratic term in Eq.~(\ref{eq7}). The total iteration number is notated as $Q$. While the gradient of the quadratic term can be easily calculated, it remains troublesome on how to obtain the gradient representation of the above prior terms. In this regard, we propose to learn the prior gradients with the network and can thus be directly learned from training data.

After gradient descent, accoding to Eqns.(\ref{eq6}) and (6), we can reconstruct the speech and noise components as
\begin{gather}
\label{eq12}
\widetilde{\mathbf{S}}^{(q+1)} = \widetilde{\mathbf{G}}_{S}^{(q+1)}\mathbf{X} + \widetilde{\mathbf{R}}_{S}^{(q+1)}, \\
\widetilde{\mathbf{N}}^{(q+1)} = \widetilde{\mathbf{G}}_{N}^{(q+1)}\mathbf{X} + \widetilde{\mathbf{R}}_{N}^{(q+1)}.
\end{gather}
\vspace{-0.7cm}
\subsection{Proposed model-driven framework}
\label{sec:proposed-model-driven-framework}
\vspace{-0.1cm}
\subsubsection{Forward stream}
\label{sec:forward-stream}\vspace{-0.1cm}
To enable the gradient update iteratively, we devise an unfolding-style framework, whose overall diagram is shown in Fig.~{\ref{fig:architecture}}(a). It has four major parts, namely feature extractor, gradient estimator, target update, and consistency layer. Given the input of noisy complex spectrum $\mathbf{X}\in\mathbb{R}^{2\times K\times L}$, it first passes through multiple 2D convolution (2D-Conv) layers with consecutive frequency downsampling operations to extract the spectral features, say, $\mathbf{F}$. In the $q$th GDM iteration, given the input $\left\{\mathbf{F}, \widehat{\mathbf{S}}^{(q-1)}, \widehat{\mathbf{N}}^{(q-1)} \right\}$, the gradient estimator will predict the prior gradients \emph{w.r.t.} the four parameters in Eqn.({\ref{eq7}}) and implement the gradient descent for updating, which are utilized to reconstruct the speech and noise targets simultaneously. The updated parameters are further modified as the inputs of the next iteration via the consistency layer~{\cite{wisdom2019differentiable}}, whose calculation process is detailed in Fig.~{\ref{fig:architecture}}(f). For practical implementation, we unfold the process for $Q$ times, and the whole forward stream is formulated as
\begin{gather}
\label{eq13}
\mathbf{F} = \text{Encoder}\left(\mathbf{X}\right),\\
\mathbf{\Omega}^{(q)} = \text{GradientUpdate}\left(\mathbf{F}, \mathbf{\Omega}^{(q-1)}\right),\\
\left\{\widetilde{\mathbf{S}}^{(q)}, \widetilde{\mathbf{N}}^{(q)}\right\} = \text{TargetUpdate}\left(\mathbf{\Omega}^{(q)}\right),\\
\left\{\widehat{\mathbf{S}}^{(q)}, \widehat{\mathbf{N}}^{(q)}\right\} = \text{ConsistencyLayer}\left\{\widetilde{\mathbf{S}}^{(q)}, \widetilde{\mathbf{N}}^{(q)}\right\},
\end{gather}
where $\mathbf{\Omega}^{(q)}$ is the parameter set for the above four parameters and $q\in\left\{1,...,Q\right\}$. Note that as decent parameter initialization is indispensable for later gradient updates, we adopt the network with the same structure as the gradient estimator to generate the initialized parameter estimation $\emph{i.e.}$, $\widetilde{\mathbf{G}}_{S}^{(0)},\widetilde{\mathbf{G}}_{N}^{(0)}, \widetilde{\mathbf{R}}_{S}^{(0)}, \widetilde{\mathbf{R}}_{N}^{(0)}$.  
\vspace{-0.2cm}
\subsubsection{Feature extractor}
\label{sec:feature-extractor}\vspace{-0.1cm}
Like~{\cite{li2022glance}}, in the feature extractor, we utilize recalibration encoding layers (RELs) to gradually downsample the feature map and abstract the features. The internal structure of each REL is shown in Fig.~{\ref{fig:architecture}}(e). Except for 2D-GLU~{\cite{dauphin2017language}}, a UNet-block is followed with residual connection~{\cite{qin2020u2}}, which is akin to UNet and takes the current feature map as the input and further encodes the feature. Compared with vanilla 2D-Conv, the UNet-block can explore features with multiple scales. Besides, the introduction of residual connection can effectively recalibrate the feature distribution and preserve the spectral patterns. 
\vspace{-0.3cm}
\subsubsection{Gradient estimator}
\label{sec:gradient-estimattor}
\vspace{-0.1cm}
The internal structure of the proposed gradient estimator (GE) is presented in Fig.~{\ref{fig:architecture}}(b). As stated above, the input includes the extracted feature $\mathbf{F}$, the modified speech $\widehat{\mathbf{S}}^{(q)}$, and noise $ \widehat{\mathbf{N}}^{(q)}$ after the consistency layer. As Fig.~{\ref{fig:architecture}}(b) shows, three gradient calculators are adopted, where the gain gradient calculator (GGC) is used to derive the gain gradients of both speech and noise, and another two complex residual gradient calculators (RGCs) aim at gradient prediction of the speech and noise residual, respectively. Note that we share the GGC for speech and noise as the gain function actually serves as the speech presence probability (SPP) in traditional SE algorithms and that of speech and noise are complementary from statistical perspective~{\cite{yoshioka2015ntt}}.

The internal structure of RGC is presented in Fig.~{\ref{fig:architecture}}(c). Take the speech branch as an example, we first flatten the complex spectrum as $\widehat{\mathbf{S}}^{(q)}\in\mathbb{R}^{2K\times L}$ and then concatenate with $\mathbf{F}$ as the network input. The network first compresses the feature with 1D-GLU and then models the gradient distribution with stacked temporal convolution networks (TCNs)~{\cite{bai2018empirical}}. To alleviate the parameter burden, we adopt the simplified version, termed as S-TCNs~{\cite{zhang2020deepmmse}}, which can dramatically decrease the parameters. The prior gradients of the complex residual are attained with two linear layers, namely for real and imaginary (RI) parts. For GGC, it has the similar structure except the input is the concatenation of $\mathbf{F}$ and the magnitude of speech and noise, say, $\text{Concat}\left( \mathbf{F}, \widehat{\mathbf{S}}^{(q)}, \widehat{\mathbf{N}}^{(q)}\right)$. The gain gradients \emph{w.r.t.} speech and noise are obtained via two linear layers.
\vspace{-0.3cm}
\subsubsection{Target fusion}
\label{sec:target-fusion}
\vspace{-0.1cm}
After repeating target updates, we obtain the estimates of speech and noise, \emph{i.e.}, $\widetilde{\mathbf{S}}^{(Q)}$, $\widetilde{\mathbf{N}}^{(Q)}$. Then another problem arises, \emph{how can we fuse the estimated speech and noise components to obtain the final speech estimation?} A common strategy is to apply a network to estimate time-frequency (T-F) bin-level weights and fuse both components dynamically~{\cite{zheng2021interactive}}, \emph{i.e.}, $\mathbf{M}\widetilde{\mathbf{S}}^{(Q)} + \left(\mathbf{1} - \mathbf{M}\right)\left(\mathbf{X}- \widetilde{\mathbf{N}}^{(Q)}\right)$. However, it is a linear combination within each T-F bin. Motivated by~{\cite{li2020two}}, we propose a residual recalibration module to better fuse the speech and noise parts. Concretely, given the original noisy and the estimated speech and noise as the input, a network is employed to estimate the residual structure, which is then added by the estimated speech:
\begin{equation}
\label{eq14}
\widetilde{\mathbf{S}}^{(Q)'} \leftarrow \widetilde{\mathbf{S}}^{(Q)} + \text{FuseNet}\left(\mathbf{X}, \widetilde{\mathbf{S}}^{(Q)}, \widetilde{\mathbf{N}}^{(Q)}\right).
\end{equation}

Differnet from the linear combination, it works with residual learning and gives nonlinear output, which can better leverage the complementarity between speech and noise.
\vspace{-0.35cm} 
\subsubsection{Loss function}
\label{sec:losss-function}
\vspace{-0.1cm} 
As the unfolding structure is utilized in the forward stream, we adopt the weighted loss for network training, given as
\begin{gather}
\label{eq15}
\mathcal{L} = \sum_{q=0}^{Q}\gamma_{q}\mathcal{L}_{q} + \zeta\mathcal{L}_{Q}^{(f)},
\end{gather}
where $\gamma_{q}$ and $\zeta$ are the weighting coefficients, and $\mathcal{L}_{Q}^{(f)}$ denotes the loss in the target fusion stage. Here $\gamma_{q}$ and $\zeta$ are empirically set to 0.1 and 1, respectively. For $\mathcal{L}_{q}$ and $\mathcal{L}_{Q}^{(f)}$ we have
\begin{equation}
\label{eq16}
\mathcal{L}_{q} =\frac{1}{2}\left( \mathcal{L}\left(\widetilde{\mathbf{S}}^{(q)\beta},  \mathbf{S}^{\beta}\right)+\mathcal{L}\left(\widetilde{\mathbf{N}}^{(q)\beta},\mathbf{N}^{\beta}\right)\right),
\end{equation}
\begin{equation}
\label{eq17}
\mathcal{L}_{Q} = \mathcal{L}\left(\widetilde{\mathbf{S}}^{(Q)'\beta},  \mathbf{S}^{\beta}\right),
\end{equation}
where the MSE criterion is adopted for network training and $\beta$ is the power-compressed coefficient and empirically set to 0.5~{\cite{li2021importance}}. Besides, the RI loss with magnitude constraint is adopted, which can mitigate the compensation effect in the complex spectrum recovery~{\cite{li2020two}}.
\vspace{-0.2cm}
\renewcommand\arraystretch{0.85}
\begin{table}[t]
	\caption{Ablation study on the proposed MDNet. The values are specified with PESQ/ESTOI(\%)/SISNR(dB) format. \textbf{BOLD} indicates the best score in each case. All the values are averaged among different SNRs and nosies in the test set.}
	\Large
	\centering
	\resizebox{\columnwidth}{!}{
		\begin{tabular}{c|ccccccc}
			\specialrule{0.1em}{0.25pt}{0.25pt}
			\multirow{2}*{Entry} &Param. &MACs  &\multirow{2}*{$Q$} &Fusion &\multirow{2}*{PESQ$\uparrow$} &\multirow{2}*{ESTOI(\%)$\uparrow$} &\multirow{2}*{SISNR(dB)$\uparrow$}\\
			&(M) &(G/s) & &type & & &\\
			\specialrule{0.1em}{0.25pt}{0.25pt}
			1a &\textbf{3.00}  &\textbf{2.16} &0 &R &2.71 &74.05 &10.00\\
			1b &4.79 &2.34  &1 &R &2.75 &74.95 &10.50\\
			1c &6.57 &2.52 &2 &R &2.76 &75.44 &10.66\\
			1d &8.36 &2.70 &3 &R &2.79 &76.03 &10.81\\
			1e &10.15 &2.88 &4 &R &2.79 &76.06 &10.78\\
			1f &11.93 &3.06 &5 &R &\textbf{2.82} &\textbf{76.55} &\textbf{10.93}\\
			1g &13.72 &3.24 &6 &R &2.79 &76.21 &10.84\\
			\specialrule{0.1em}{0.25pt}{0.25pt}
			2a &7.85 &2.38 &3 &A &2.73 &74.41 &10.38\\
			2b &8.33 &2.68  &3 &G &2.77 &75.30 &10.59\\
			\specialrule{0.1em}{0.25pt}{0.25pt}
	\end{tabular}}
	\label{tbl:ablation-studies}
	\vspace{-0.6cm}
\end{table}
\renewcommand\arraystretch{0.75}
\begin{table*}[t]
	\setcounter{table}{2}
	\caption{Quantitative comparisons with other state-of-the-art systems on the DNS Challenge dataset. ``-'' denotes no published result.}
	\normalsize
	\centering
	\scalebox{0.85}{
		\begin{tabular}{cccccccccccc}
			\specialrule{0.1em}{0.25pt}{0.25pt}
			\multirow{2}*{Methods} &\multirow{2}*{Year} &\multirow{2}*{Do.}  &\multicolumn{4}{c}{w/ Reverberation} & \multicolumn{4}{c}{w/o Reverberation}\\
			\cmidrule(lr){4-7}\cmidrule(lr){8-11}
			& &  &WB-PESQ$\uparrow$ &PESQ$\uparrow$ &STOI(\%)$\uparrow$ &SISNR (dB)$\uparrow$ &WB-PESQ$\uparrow$ &PESQ$\uparrow$ &STOI(\%)$\uparrow$ &SISNR(dB)$\uparrow$\\
			\specialrule{0.1em}{0.25pt}{0.25pt}
			Noisy  &- &- &1.82 &2.75 &86.62 &9.03 &1.58 &2.45 &91.52 &9.07\\
			NSNet~{\cite{reddy2020interspeech}} &2020 &T-F &2.37 &3.08 &90.43 &14.72 &2.15 &2.87  &94.47 &15.61\\
			DTLN~{\cite{westhausen2020dual}}  &2020 &T-F &- &2.70 &84.68 &10.53 &- &3.04 &94.76  &16.34\\
			DCCRN~{\cite{hu2020dccrn}} &2020 &T-F &- &3.32 &- &- &- &3.27 &- &- \\
			FullSubNet~{\cite{hao2021fullsubnet}} &2021  &T-F &2.97 &3.47 &92.62 &15.75 &2.78  &3.31 &96.11 &17.29\\
			TRU-Net~{\cite{choi2021real}} &2021 &T-F &2.74 &3.35 &91.29 &14.87 &2.86  &3.36 &96.32 &17.55\\
			CTS-Net~{\cite{li2020two}} &2021 &T-F &3.02 &3.47 &92.70 &15.58 &2.94 &3.42  &96.66  &17.99\\
			GaGNet~{\cite{li2022glance}} &2022 &T-F &3.18 &3.57 &93.22 &16.57 &3.17 &\textbf{3.56} &97.13 &18.91 \\
			\textbf{MDNet(Ours)} &2022 &T-F &\textbf{3.24} &\textbf{3.59} &\textbf{93.61} &\textbf{16.94} &\textbf{3.18} &\textbf{3.56} &\textbf{97.20} &\textbf{19.17}\\
			\specialrule{0.1em}{0.25pt}{0.25pt}
	\end{tabular}}
	\label{tbl:dns1}
	\vspace{-0.5cm}
\end{table*}
\vspace{-0.2cm}
\section{Experimental setup}
\label{sec:experiments-setup}
\vspace{-0.2cm}
Two datasets are adopted to carry out the experiments. WSJ0-SI84~{\cite{paul1992design}} consists of 7138 utterances by 83 speakers (42 males and 41 females). 5428 and 957 utterances are selected for training and validation, and 150 utterances spoken by unseen speakers are for testing. Around 20,000 types of noises are randomly selected from the DNS-Challenge noise set as the training noise set and mixed with clean utterances under SNRs $[-5\rm{dB}, 0\rm{dB}]$ with $1\rm{dB}$ interval. For testing, three challenging unseen noises are chosen, namely babble, factory1 from NOISEX92~{\cite{varga1993assessment}} and cafeteria from CHiME-3~{\cite{barker2015third}} under three SNRs, \emph{i.e.}, $\left\{-3, 0, 3\right\}\rm{dB}$. Totally, we create 150,000, 10,000 noisy-clean pairs for training and validation. For testing, 150 pairs are created for each SNR. Interspeech2020 DNS-Challenge\footnote{https://github.com/microsoft/DNS-Challenge} provides 562.72 hours clips from 2150 speakers and 181 hours of 60,000 clips from 150 classes. For model evaluation, it provides a non-blind validation set with two categories, namely with and without reverberation, and each includes 150 noisy-clean pairs. Following the script given by the organizer, we create around 3000 hours of pairs for training, and the input SNR ranges from -5\rm{dB} to 15\rm{dB}.

In the feature extractor, the kernel size and stride of 2D-Convs are $\left(1, 3\right)$ and $\left(1, 2\right)$ along the time and frequency axes, respectively. For each UNet-block, the kernel size is $\left(2, 3\right)$.  The channel number of 2D-Convs remains 64 by default. Let us notate the number of (de)encoder layers in the $i$th UNet-block as $U_{i}$, then $U = \left\{4,3,2,1,0\right\}$ where $0$ means no UNet-block is used. Two groups of S-TCNs are employed, each of which includes four temporal convolution modules (TCMs), with the kernel size of the dilated Convs and dilation rates being 3 and $\left\{1,2,5,9\right\}$, respectively. Note that we take causal convolution operations by zero-padding along the past frames. The optimization step $\mathbf{\eta}$ is set as trainable and initialized at 0.01.

All the utterances are sampled at 16 kHz. 20 ms Hanning window is utilized with 50\% overlap between adjacent frames. 320-point FFT is utilized, leading to 161-D features in the frequency axis. The model is trained on the Pytorch platform with an NVIDIA V100 GPU. Adam optimizer is adopted for network training ($\beta_{1}=0.9$, $\beta_{2}=0.999$) and the total epoch is 60 with batch size being 8. The learning rate is initialized at 5e-4 and we halve the value if the validation loss does not decrease for consecutive two epochs. For fusing speech and noise components, we adopt the same ``Encoder-TCN-Decoder'' structure as~{\cite{li2020two}} except we adopt a lightweight version by halving the channel number in the 2D-Convs.
\renewcommand\arraystretch{0.85}
\begin{table}[t]
	\setcounter{table}{1}
	\caption{Quantitative comparisons with other SOTA systems on WSJ0-SI84 dataset. Scores are averaged upon different testing cases. ``Do.'' denotes the tranform domain of the method.}
	\Large
	\centering
	\resizebox{\columnwidth}{!}{
		\begin{tabular}{cccccccc}
			\specialrule{0.1em}{0.25pt}{0.25pt}
			\multirow{2}*{Methods} &\multirow{2}*{Year}  &\multirow{2}*{Do.} &Param. &MACs &\multirow{2}*{PESQ$\uparrow$}  &ESTOI$\uparrow$ &SISNR$\uparrow$\\
			& & &(M) &(G/s) & &(\%) &(dB) \\
			\specialrule{0.1em}{0.25pt}{0.25pt}
			Noisy  &- &- &- &- &1.82 &41.97 &0.00\\
			DDAEC~{\cite{pandey2020densely}} &2020 &T &4.82 &36.56 &2.76 &74.84 &10.85 \\
			DEMUCAS~{\cite{defossez2020real}} &2020  &T &18.87 &4.35 &2.67 &76.23 &11.08\\
			GCRN~{\cite{tan2019learning}} &2020 &T-F &9.77 &2.42 &2.48 &70.68 &9.21 \\
			DCCRN~{\cite{hu2020dccrn}} &2020 &T-F&\textbf{3.67} &11.13 &2.54 &70.58 &9.47 \\
			PHASEN~{\cite{yin2020phasen}} &2020 &T-F&8.76 &6.12 &2.73 &71.77 &9.38\\
			FullSubNet~{\cite{hao2021fullsubnet}} &2021 &T-F &5.64 &31.35 &2.55 &65.89 &9.16\\
			CTSNet~{\cite{li2020two}} &2021 &T-F &4.35 &5.57 &2.86 &76.15 &10.92 \\
			GaGNet~{\cite{li2022glance}} &2022 &T-F &5.94 &\textbf{1.63} &2.86 &76.87 &10.93\\
			\textbf{MDNet(Ours)} &2022 &T-F &8.36 &2.70 &\textbf{2.88} &\textbf{77.37} &\textbf{11.12} \\
			\specialrule{0.1em}{0.25pt}{0.25pt}
	\end{tabular}}
	\label{tbl:wsj0-si84-result}
	\vspace{-0.6cm}
\end{table}
\renewcommand\arraystretch{0.85}
\vspace{-0.4cm}
\section{Results and analysis}
\label{sec:results-and-analysis}
\vspace{-0.2cm}
\subsection{Ablation study}
\label{sec:ablation-study}
\vspace{-0.2cm}
Around 100 hours of training data from the WSJ0-SI84 corpus are used to conduct the ablation study in terms of step number $Q$ and fusion type. Three evaluation metrics are utilized, namely PESQ~{\cite{rix2001perceptual}}, ESTOI~{\cite{jensen2016algorithm}}, and SISNR~{\cite{le2019sdr}. Higher values indicate better performance. Quantitative results are shown in Table~{\ref{tbl:ablation-studies}} and several observations can be made. First, with the increase of steps, one can observe notable performance improvements over the non-update case (entry 1a), which shows that by gradient updating, we can obtain more accurate parameter estimation, leading to better reconstruction results. However, with more steps, the performance tends to get saturated and even degraded, \emph{e.g.}, from 1f to 1g. It can be explained as the estimation with GDM tends to get converged and GDM can not guarantee consistent optimization with the increase of steps for the non-convex optimization problem. Second, we compare three fusion strategies, namely ``R'' (the proposed fusion method), ``G'' (dynamic weighting), and ``A'' (average), and one can see that our method yields the best performance among above fusion schemes. It reveals the superiority of the proposed nonlinear residual fusion over dynamic weighting. Note that the average operation is the special case of dynamic weighting, where the weighting coefficients remain 0.5 for all T-F bins. However, as the local SNR varies a lot in different frequency bands, it is not reasonable to combine both parts with fixed weights, which can be proved from the notable performance degradations from entry 2b to 2a.
\vspace{-0.4cm}
\subsection{Comparsion with state-of-the-art methods}
\label{sec:comparison-with-sota-methods}
\vspace{-0.2cm}
Based on the analysis in the ablation study, we choose entry 1d as the default network configuration, which well balances between calculation complexity and performance, to compare with current top-performed SE systems. Quantitative results on the WSJ0-SI84 dataset are presented in Table~{\ref{tbl:wsj0-si84-result}}. Compared with another eight baselines, the proposed approach yields the highest scores among three objective metrics, validating the superiority of our method in speech quality and intelligibility. Note that despite our method has around 8.36M trainable parameters, it is rather advantageous in MACs, say, 2.70G/s, which reveals the gap in trainable parameters and computation complexity.

We also report the quantitative results on the DNS-Challenge non-blind test set, as shown in Table~{\ref{tbl:dns1}}. The wide-band version of PESQ (WB-PESQ)~{\cite{itu862}} and STOI~{\cite{taal2010short}} are also listed for evaluation. From the results, one can see that again, our method achieves the highest scores in different metrics over previous top-performed systems, which further attests to the superiority of our method under both reverberant and anechoic environments. Remark that different from previous literature where the mapping process lacks adequate interpretability, our method follows the MAP criterion and is explicitly optimized with the gradient descent method. Therefore, we think it is a promising direction to leverage the advantage of model-based methods to gradually open the black-box of DNNs in the speech enhancement area.
\vspace{-0.35cm}
\section{Conclusions}
\label{sec:conclusion}
\vspace{-0.2cm}
In this paper, we propose a model-driven approach to tackle single-channel speech enhancement. Specifically, based on the maximum a posteriori criterion, the original problem is formulated into the joint posterior estimation of speech and noise, and it is proposed that deep prior distribution is learned via the network from training data. The framework is devised with the unfolding structure and the gradient descent method is employed to update parameter estimation and facilitate the target reconstruction progressively. Besides, another network serves as the fusion module to further recover the speech component from previous estimations. Experimental results on the WSJ0-SI84 and DNS-Challenge datasets show that the proposed approach performs favorably against previous top-performed SE systems.

\vfill\pagebreak
\bibliographystyle{IEEEtran}
\bibliography{refs}

\end{document}